 \documentclass[aps,preprint,superscriptaddress,groupedaddress]{revtex4}  
\usepackage{graphicx}  
\usepackage{dcolumn}   
\usepackage{bm}        
\usepackage{amssymb}   
\usepackage{amsmath}
\usepackage{dsfont}

\newcommand{\be}{\begin{equation}}
\newcommand{\ee}{\end{equation}}
\newcommand{\bea}{\begin{eqnarray}}
\newcommand{\eea}{\end{eqnarray}}
\newcommand{\bee}{\begin{eqnarray*}}
\newcommand{\eee}{\end{eqnarray*}}

\newcommand{\nnu}{\nonumber\\}

\newcommand{\mt}{\tilde m}

\newcommand{\at}{\tilde a}

\def\mt{{\ifmmode\td M_t\else $\td M_t$\fi}}
\def\as{{\ifmmode\alpha_s\else$\alpha_s$\fi}}

\let\lam=\lambda
\let\td=\tilde

\def\co#1{{\ifmmode{\cal O}_{#1}\else${\cal O}_{#1}$\fi}}
\def\cs#1{{\ifmmode{\cal S}_{#1}\else${\cal S}_{#1}$\fi}}
\def\at{{\ifmmode{\tilde A}\else$\tilde A$\fi}}

\def\fr#1.#2.{{#1\over #2}}

\def\mg{{\ifmmode M_{GUT}\else $M_{GUT}$\fi}}

\newcommand{\oot}{\overline {126}}

\newcommand{\ovl}{\overline}
\newcommand{\boot}{${\bf{\oot}}$ }

\begin{document}




\title{  Bajc-Melfo  Vacua enable Yukawon ultraminimal grand unified theories }
\author{ Charanjit S. Aulakh\footnote{aulakh@pu.ac.in; aulakh@iisermohali.ac.in;
 http://14.139.227.202/Faculty/aulakh/}  }

\affiliation{Indian Institute of Science Education and Research
Mohali,\\ Sector 81, S. A. S. Nagar, Manauli PO 140306, India }

\date{\today}

  \begin{abstract}
 Bajc-Melfo(\textbf{BM})   two field  ($S,\phi$)  superpotentials
   define metastable F-term supersymmetry breaking   vacua
  suitable as    hidden sectors    for calculable and realistic
  family and Grand Unification  unification models.
The undetermined vev $<S_s>$ of the Polonyi field
 that  breaks Supersymmetry  can be fixed either
 by coupling to N=1 Supergravity or by radiative corrections.
 \textbf{BM}   hidden sectors extend   to
   symmetric multiplets $(S,\phi)_{ab}$ of a gauged  $O(N_g)$
   family symmetry, broken at the GUT scale,  so that the $O(N_g)$
   charged component vevs   $<\hat S_{ab}>$ are  also
     undetermined    before  accounting for the $O(N_g)$ D-terms:
 which fix them by  cancellation against     D-term contributions
      from the visible sector.  This  facilitates Yukawon Ultra
      Minimal GUTs(YUMGUTs) proposed in [C.S.Aulakh and C.K.Khosa, Phys.Rev.D 90,045008(2014)] by   relieving
 the visible sector from the need to give
  null  D-terms for the family symmetry  $ O(N_g)$.
We analyze  symmetry breaking and and spectra of the hidden sector
fields in the Supergravity resolved case  when  $N_g=1,2,3$.
Besides the Polonyi field $S_s$, most of the superfields $\hat
S_{ab}$ remain light, with fermions getting masses only from loop
corrections. Such  modes  may yield novel dark matter lighter than
100 GeV.  Possible  Polonyi and moduli problems associated with
the  the fields  $S_{ab}$ call  for   detailed
 investigation of loop effects due to the Yukawa and gauge
 interactions in the hidden sector and of post-inflationary
  field relaxation dynamics.

\end{abstract}


 \maketitle

%

\section{Introduction}

Supersymmetry imposes remarkable restrictions on its own
spontaneous violation and thus   severely constrains   models of
BSM physics.
Traditionally\cite{cremmer,ohta,nath,weinberg82,halllykkwein}
gravity mediation has been used to generate phenomenologically
acceptable soft Susy breaking terms for globally supersymmetric
GUTs. This requires supersymmetry breaking in a hidden sector with
no superpotential couplings to the `visible' sector in which we
live. The GUT Minkowski space vacuum is
metastable\cite{weinberg82} in the sense of there being a Susy
preserving vacuum with lower (negative) energy, but it is thought
that it cannot decay to the lower vacuum by tunnelling; or at any
rate the timescale for doing so is much larger than the age of the
universe.  The vacuum in the gravity mediated scenario is defined
as the perturbation of a Global Susy preserving minimum by a
hidden sector with a \emph{global} minimum which breaks
supersymmetry spontaneously due to specialized
superpotentials(Polonyi,O'Raifeartaigh
 \cite{polo,raff} etc) or due to  strong
dynamics(gaugino condensation etc). After long struggles to tame
the hard problem of finding phenomenologically usable  Susy
breaking global minima it was
re-emphasized\cite{dimodvalrattgiu,intrillshih} that metastable
i.e local minima could serve just as well as global minima, since
  the unifying theory would anyway  have vacua at
lower energy\cite{weinberg82} than the phenomenologically relevant
Minkowski vacuum on which the MSSM lived.   This realization was
liberating for phenomenologists because  models defined on
supersymmetry breaking vacua which are  only local minima offer  a
much larger range of candidate vacua. Various strongly coupled
supersymmetric gauge theories \cite{intrillshih,Lutyagashe}were
studied to provide the Susy breaking dynamics albeit now
considering also metastable vacua. Unfortunately even the
metastable dynamical Susy breaking models seem  so arbitrarily
elaborate as to rival the complexity of the phenomenological
theory (GUT,MSSM etc) for which they were supposedly to provide
the service of Supersymmetry breaking: making testability of both
a moot issue.

Recently Bajc and Melfo (BM)\cite{BM} suggested that,  from a
`calculable unification'  viewpoint, it would be more interesting
to consider Susy breaking metastable vacua of simple two field
hidden sector superpotentials. BM models  are related to the
classic O'Raifeartaigh susy breaking  superpotentials and are
variants of a class of models with   susy breaking local minima
studied systematically earlier  in \cite{ray}. Such
superpotentials typically have flat directions and there is a
longstanding  vision whereby the determination of such flat
directions(running out of susy breaking minima) by radiative
corrections\cite{wittenhilo} or supergravity\cite{ovrab,csahilo}
might profitably fix the undetermined vacuum expectation
values(VEVs) at a high Unification scale. Bajc and Melfo took up
this challenge and formulated models where the  undetermined VEVs
were those of \textbf{24}  or \textbf{75} irreps of an $SU(5)$
unified theory and the flat directions were determined by
radiative corrections. The use of of N=1 Supergravity potentials
to lift the flat directions rolling out of the Bajc-Melfo   Susy
breaking local minimum has  not   been  analyzed.

From a Grand Unified viewpoint the only credible sign of flavour
unification so far so far is  third family  Yukawa
 unification at large $\tan \beta$ in SO(10)   at  scales of order
  $M_X > 10^{16}$ GeV. On the other hand   minimal
  Supersymmetric SO(10)  GUTs\cite{aulmoh,ckn,abmsv,bmsv,ag2,
gmblm,blmdm,nmsgut}  have been shown to be completely realistic
and even to suppress $d=5$ proton decay\cite{gutupend,bstabhedge}
by a novel and generic mechanism based on the necessary  emergence
of a pair of light doublet Higgs in the effective MSSM. Motivated
by these two clues  we exploited  the defining(i.e  \boot )
representation of MSGUTs(which gets large vevs yet couples to
matter fermions)  to design a \emph{renormalizable}  Grand Unified
Yukawon/familion model\cite{yukawon} in which we generate
hierarchical matter fermion  Yukawas from
  flavour bland  parameters of the MSGUT extended by an $O(N_g)$
   family symmetry. Family gauge symmetry is  broken at the unification
 scale and the visible sector GUT Higgs fields  effloresce  into
 symmetric ($\mathbf{210,126,\oot,10}$)  or anti-symmetric
 ($\mathbf{120}$ ) irreps of the family group (as dictated by the
 properties of bilinears in the  \textbf{16}-plet matter  irreps in SO(10)
  GUTs), while the matter \textbf{16-}plet Yukawa couplings become
   flavour bland.  The neat
and exactly soluble\cite{abmsv} gauge symmetry breaking
($SO(10)\rightarrow \rm{MSSM}$) of the family singlet MSGUT Higgs
system then generalizes straightforwardly for the family symmetry
variant MSGUT Higgs multiplets. The supersymmetric   breaking of
family symmetry will be discussed in detail in this paper.
Extraction of the light MSSM Higgs from the plethora of MSSM
doublets present in the $O(N_g)\times SO(10)$ GUT  via calculation
of the `Higgs fractions'\cite{abmsv,ag1,ag2,nmsgut} (which specify
the make up of the MSSM Higgs) determined by flavour blind GUT
parameters generates candidate  flavour structures.

In implementing this flavour generation scenario an obstacle
arises. Although the solution of the F and D term conditions for
the GUT multiplets in the SO(10) sector with flavour  non-diagonal
VEVs is straightforward, it is very hard or impossible to ensure
that the same configurations simultaneously cancel the family
symmetry D-terms. It thus becomes necessary to introduce
additional fields to soak up the the contribution   of the GUT
VEVs   to the $O(N_g)$ D terms. Since minimality, simplicity and
solubility of the GUT Higgs sector symmetry breaking is a central
virtue of MSGUTs   it is natural to ask whether the  additional
fields might not be located in a hidden sector with its own flat
directions arising from supersymmetry breaking in a way which did
not interfere with the already accomplished MSGUT symmetry
breaking\cite{aulmoh,ckn,abmsv}. Besides making a completely
unexpected connection between family symmetry  and supersymmetry
breaking this idea is straightforward to implement in a context
where the BM flat directions are lifted by a combination of
supergravity and $O(N_g)$ gauge D term minimization effects.

 A notable phenomenological implication of our model  is the presence
  of  weakly coupled light fields(`moduli').
In particular, in the   supergravity mediated scenario presented
here there is a Polonyi (scalar) field (Planck scale vev, weak
scale mass and intermediate scale susy breaking F-term). In
addition there are other light fermions and scalars  associated
with the flat directions of the family non singlet BM fields.
These modes may be interesting light( $<50$ GeV) Susy WIMP dark
matter  of the sort perhaps indicated by the DAMA/LIBRA experiment
\cite{damalibra} and so far completely missing from MSGUT spectra.
On the other hand this clutch of modes may also  pose the typical
problems associated with `moduli' fields found in other
fundamental theories\cite{modulicon}.
   Recall that the Polonyi problem
 arises  due to late decay of the oscillations
  of the supersymmetry breaking  Polonyi field while it settles into the zero temperature
  potential's    minimum because of the suppression of   its couplings by the Planck scale.
 The dark matter candidates' role  in cosmogony also needs to
 be checked for consistency. In view of the new Yukawa and gauge
  interactions of the hidden sector the cosmological implications
  require a separate detailed study and thus  should not prematurely prejudice the
  novel flavour generation aspects of our model.
  In this paper we give the relevant details for the symmetry breaking, VEVs and
masses in the BM sector  for direct use in our `yukawonification'
\cite{yukawon} models. We will return to a study of cosmological
constraints and alternative models elsewhere.

\section{The Bajc Melfo Vacua and Supergravity}

The Bajc-Melfo superpotential\cite{BM} is a quadratic variant of
of the the O'Raifeartaigh superpotential and its relatives
\cite{raff,ray} which breaks superymmetry at a {\it{metastable}}
minimum
 \bea W_{BM}(S_s,\phi_s)= W_0+ S_s(\mu_B \phi_s + \lambda_B
\phi_s^2)\eea

We have added a constant term $W_0$ for later convenience.  It is
then easy to see\cite{ray,BM} that besides the Susy preserving
global minima at $S_s=\phi_s=0$ and    $ S_s=0,
\phi_s=-\mu_B/\lambda_B $
 with zero vacuum energy,  there is also a local minimum at
$<\phi_s>=-\mu_B/2 \lambda_B$ with $<S_s>$ undetermined provided
\bea |<S_s>|\ge |{\frac{<\phi_s>}{\sqrt{2}}}|\label{BMW}\eea This
local minimum breaks supersymmetry since $<F_S>=-\mu_B^2/4
\lambda_B\equiv \theta$. Thus the   chiral fermion in the
multiplet $S_s$ provides the  goldstino field while  its partner
scalar, whose VEV remains undetermined, is massless. When
supersymmetry is made local by coupling the theory  to gravity the
goldstino can be gauged away by a local supersymmetry
transformation leaving   a massive gravitino while the scalar
picks up the gravitino mass common to  light scalars in gravity
mediation. For clarity we first review  the effect of coupling to
supergravity on extrema of globally supersymmetric
Lagrangians\cite{weinberg82}.

The generic $N=1$ supergravity
potential\cite{cremmer,nath,halllykkwein}
  for scalar fields $z_I$ interacting through a superpotential
  $W(z_I)$ ( with canonical Kahler potential $K(z,\bar z)$ and gauge Kinetic
  function $f_{\alpha\beta}(z)=\delta_{\alpha\beta} $, $\kappa$ is
  the Planck Length)
 \bea V&=& E (\sum_I|{\cal{F}}^I|^2 - 3
\kappa^2 |W|^2)   +  \sum_\alpha  {\frac{g^2_\alpha}{4}}
{\big{[}}{\frac{({\cal{F}}^I (T^{\alpha})_I^J z_J)^2}{ \kappa^4
W^2 }} + h.c. {\big{]}} \nonumber
 \nnu {\cal{F}}^I &=& {\frac{\partial W}{\partial z_I}} +
  {z^*}^I \kappa^2 W \qquad;\qquad E \equiv e^{\kappa^2
\sum_I |z_I|^2}\label{Vsgrav}\eea

  As is well known, the constant term
in the superpotential $W$, which is irrelevant in global
supersymmetry,  can be used to tune the vacuum energy in
\emph{one} of its extrema to zero after coupling to Supergravity.
This means that for a  supersymmetric minimum  with vevs
$<z_I>=\bar z_I $ derived from  \bea   <F^I>=<\partial W/\partial
z_I>=0 \quad;\quad <D^\alpha>= <z^{*I} T^{\alpha J}_I z_J
>=0 \eea there is no shift in the global vevs $<\bar{z}_I>$ at
all: since by tuning $<W>={\overline{W}}=0$ we ensure that the vev
$\overline{\cal{F}}^I=\overline{ {F}}^I=0$, the vacuum energy at
the old minimum  vanishes and so do terms in the extremization
condition which come from the variations of the fields in the
exponential factor (since they multiply a separately vanishing
factor). It is however amusing that the D-term vev in
(\ref{Vsgrav}) above is then indeterminate : seemingly indicating
and incompatibility between Supergravity and preserved
supersymmetry with$<W>=0$ !

On the other hand if the global superpotential breaks
supersymmetry spontaneously(as for example, the Polonyi or
O'Raifeartaigh superpotentials  at a global minimum or the BM
superpotential at a local minimum), then the non vanishing of $F$
at the minimum means that the total superpotential vev must be
nonzero to cancel the global contribution to the  vacuum energy in
eqn(\ref{Vsgrav}). The vevs of the supersymmetry preserving
visible sector fields may also shift from their globally
determined values by small terms of order $O(m_{3/2})$ and
maintain zero value for their local F-terms. Special behaviour is
seen only for the components of the sliding multiplet $S$ which
breaks supersymmetry.

Using the potential (\ref{Vsgrav}) in  the pure  BM case we see
that the field $S_s$ whose value was \emph{undetermined } at the
local/metastable minimum of the global superpotential described
above  now picks up a potential which fixes its value to be of
order the Planck scale. The other BM field $\phi_s$ which already
had a large  vev (think of it as  $\bar\phi_s \sim 10^{12}$  GeV
on phenomenological grounds in the context of the gravity
mediated(SUGRY)  models we are developing) will only experience a
tiny shift in its vev. To leading order we can determine the vev
$<S_s>\equiv \bar S_s$ by fixing $\phi_s$ at its globally
determined value $\bar\phi_s=-\mu_B/2\lambda_B$. The potential to
be minimized is then

\bea V(S_s)=e^{\kappa^2 |S_s|^2   + \delta } \{(\delta \kappa^2 |
W_0+ S_s \theta|^2 + |\theta + \kappa^2 S_s^*(W_0+ S_s \theta)|^2
- 3 \kappa^2 |W_0+ S_s \theta|^2 \}\eea

Here $\delta =\kappa^2(|\bar{\phi}_s|^2 )$ is a tiny background
fixed \emph{parameter} ($\sim 10^{-12}$) for the purposes of the
extremization w.r.t. the globally undetermined vev of $S_s$ and
has essentially no effect on the value of $S_s$ at the extremum
found. Defining dimension less variables \bea x
&=&\kappa {\frac{\widehat{W}_0}{\theta}} \quad ; \qquad y =\kappa S_s \nonumber\\
\varphi_x &=& Arg[x]   \quad ; \qquad  \varphi_y=Arg[y]\eea
 the potential becomes \bea {\widetilde
V}&\equiv&{\frac{V}{|\theta|^2}} = {\big\{}(|x|^2
+|y|^2)(\delta-3) + (1+|y|^2)^2 + \nnu && |x|^2 |y|^2 + 2
\cos(\varphi_y-\varphi_x)|x||y|(|y|^2 +\delta-2)
{\big\}}\label{dimlessV}\eea

We emphasize that $x$ (which is  the dimensionless form of the
constant term in the superpotential) is not a dynamical variable
but a parameter to be tuned  to maintain the Minkowski Vacuum
energy at zero. Similarly $\delta$ is a parameter fixed by the
globally determined background to the $S_s$ extremization  at a
lower order of the expansion in powers of $\kappa$ and not a
variable.

The only stable minimum of (\ref{dimlessV})  with respect to $y$
(i.e $S_s$)  with zero energy i.e for(field subscripts denote
partial derivatives) : \bea V=V_{|y|}=V_{ \theta_y}= 0=V_{|y|,
\theta_y}\qquad ; \qquad V_{|y|,|y|},V_{\theta_y \theta_y}
>0 \eea
is achieved as \bea   \varphi_y &=& \varphi_x   \quad ;\qquad x= 2
- \sqrt{3 -\delta} \quad ;\qquad  y= y_0= \sqrt{3-\delta} -1 \nnu
{\overline{\partial_{|y|} \partial_{|y|} \tilde{V} }} &=& 4
\sqrt{3-\delta} \quad ; \qquad {\overline{\partial_{\varphi_y }
\partial_{\varphi_y }\tilde{V} }}  = 4 \delta \sqrt{3-\delta } -16
\delta -32 \sqrt{3-\delta }+56\quad \label{hidVEV}\eea It is clear
that the condition for a local minimum  (\ref{BMW}) is satisfied
for \bea \delta < 3-
(1+\sqrt{|\frac{\kappa^2\theta}{2\lambda_B}|})^2 \simeq 2
\label{deltacond} \eea
 As mentioned $\delta $ is tiny and this condition is trivially satisfied.
 Thus we have fixed the leading contribution  $\bar S_s$ at the Planck scale by
 the combination of a gravitino scale ($m_{3/2}=\kappa^2<W>$) mass
  and a tiny ($\sim(m_{3/2}/M_{Planck})^2$) quartic coupling for the sliding
  field $S_s$. The leading order vevs ${\bar{\phi_s},\bar S_s}$
    specify a Minkowski vacuum, with the degeneracy in $S_s$ lifted, to leading order in $
\kappa\bar\phi_s \sim 10^{-13}$. There is no immediate reason to
compute next order shifts by perturbing around these vevs, but it
is straightforward to do so iteratively.

 We now wish to build upon these complementary  simple cases to first analyze  a
 model in which a hidden sector BM superpotential
$W_{BM}(S_s,\phi_s)=W_H$ with gauge singlets  is coupled to a
visible(GUT) sector with(fields $z_i$)  which has  a globally
supersymmetric minimum associated with superpotential $W_O(z_i)$
and is invariant under  a grand unifying gauge group broken
spontaneously by the vevs $\bar z_i$ at a superheavy scale
($M_G\sim 10^{16}$ GeV ) while preserving
Susy\cite{ag2,nmsgut,yukawon}. There
 is no coupling between the fields of the two sectors in the superpotential.
 So their global extrema and the non determination of $\bar S_s$ are unchanged.
 It is  eminently reasonable to pursue the determination of the
 leading correction by setting  fields $\phi,z_i$ already determined at the global
 level to those values ($\bar\phi,\bar z_i$) and determine the undetermined field i.e
 $S_s$ by minimizing (\ref{Vsgrav}) with respect to $S_s$, tuning $W_0$ to maintain
 zero energy at the vacuum.  In this case the entire discussion goes through verbatim
 with two superficial modifications :

 \begin{enumerate}
 \item The constant term in the superpotential now also receives a
 contribution from the vev of the visible sector i.e $W_O(\bar
 z_i)$.
 \item The tiny background determined  parameter  $\delta$ now also contains the visible
 sector values  :
 \bea \delta &=& \kappa^2(|\bar \phi_s|^2 + \sum_i|{\bar z_i}|^2)
 \eea
\end{enumerate}

 Thus the determination of $S_s$ is unchanged. The value of
 $\delta$ is still very small since $M_{GUT}<<M_{Planck}$, $\delta<
 10^{-6} $.   Condition (\ref{deltacond}) is still obviously
 trivially satisfied.

 Finally we arrive at the novel construction proposed in\cite{yukawon}
 where a (family) gauge symmetry links the BM hidden sector to
the visible sector,via the D terms,  even at the global level but
there is still no link at the level of the superpotential. As we
will see in the following sections   besides the field $S_s$ also
the gauge non-singlet parts $\hat S$ of the sliding field remain
undetermined by the global  F-term conditions.  \emph{However} the
visible sector contributions to the  family gauge group  D terms
\emph{will not}\cite{yukawon}  vanish. In the BM sector vevs of
the family non singlet part of $\phi$ field have zero vev so that
$\phi$ makes no contribution to the family $D-$terms. While the
undetermined $\hat S$ non-singlet BM vevs will contribute to the-
positive definite- family group D terms. Since the visible
contribution is $\sim M_{GUT}^4$ , it is obviously highly
favorable for the $\hat S$ fields to take vevs and cancel this
positive definite contribution to the global susy potential. For
definiteness let the Family gauge group be the $O(N_g), N_g=2,3$
actually used in \cite{yukawon} (for cogent structural and
phenomenological reasons). Then  an $O(N_g)$ gauge transformation
can always align the    family D-term vev with any one generator
of the gauge group and thus requires (in that aligned basis)  only
one $\hat S$ component to get a vev. It is clear that there will
be a pseudo-goldstone degeneracy among the the remaining $\hat S$
fields which may be lifted by Supergravity effects. Since field
$\hat S$ gets a vev $\bar {\hat S} <<M_{Planck}$ it too affects
the $S_s$ vev (like $\bar\phi_s,\bar z_i$) only very weakly
through $\delta$ which now takes the value
  \bea \delta &=& \kappa^2(|\bar \phi_s|^2 + \sum_i|{\bar
z_i}|^2  +Tr\bar{\hat S}^\dagger  \bar{\hat S})
 \eea

and the rest of the determination of the BM-Polonyi field $S_s$
goes through unchanged.

 As usual in gravity mediated
scenarios,  the chiral fermion in the local supersymmetry breaking
S multiplet is the global Susy breaking Goldstino that furnishes
the longitudinal mode of the gravitino which gets the mass \bea
m_{\frac{3}{2}} &=& \kappa^2 |\sqrt{\bar{E}}(W_0  +W_O(\bar z_i)+
{\ovl{W_H}}) | \eea Since this mass should be $\sim 10^3 - 10^5$
GeV in typical gravity mediated scenarios  and
  typically $ \ovl{W}_{GUT}  \sim M_X^3 > 10^{48}$ GeV, it is clear
 that $W_0$ must be used to cancel $\ovl W_{GUT}$  so that \bea
   |<W>|\simeq M_p|\theta| \simeq 10^{39}-10^{41}
 GeV^3\eea The scale of supersymmetry breaking is fixed by the
 BM  superpotential
 parameters :\bea  \sqrt{|F^S|}&=&\sqrt{|\theta|}=|\frac{\mu_B}{  2\sqrt{\lambda_B}}|\sim
 10^{10.5}-10^{11.5} \rm{ GeV} \eea  Provided $|\lambda_B | > 10^{-14}$  one
 has $|\lambda_B S| >>|\mu_B|$.

Having discussed the case $N_g=1$ in this section and indicated
  the general structure and  solution strategy for our
model\cite{yukawon}, where family gauge symmetry - but not
superpotential couplings - tie the visible and hidden sectors
together, we will discuss the details of the global
superpotentials and family representations in the next section for
$N_g=2,3$ while the moduli mass spectra associated with the $\hat
S$ ``familion moduli'' are discussed in Section 4.

A slightly subtle point that follows from the role of ${\cal{F}}$
as a local supersymmetry order  parameter is that   ${z_i,\phi_s}$
get small corrections of order $m_{3/2}$ so as to   maintain
${\cal{F}}_{z_i,\phi_s}\equiv 0$ and similarly the family gauge
non singlet  BM fields $\hat \phi$ remain zero (see the next
section) so that ${\cal{F}}_{\hat \phi}\equiv 0$. However since
$\bar S_s\sim M_{Planck},\bar{\hat S}\sim M_{GUT}$ while
$\ovl{W}\neq 0$, it is clear that $ {\cal{F}}_{S_s,\hat S}$ are
not zero but ${\cal{F}}_{\hat S}/ {\cal{F}}_{S_s} \sim
M_{GUT}/M_{Planck} <<1 $, so that $S_s$ still dominates the
supersymmetry breaking.

\section{Gauged $O(N_g)$ and   BM vacua}

 Assuming that the visible sector supersymmetry conditions are
 satisfied  the vanishing of the \emph{family symmetry} D terms  for the GUT
sector VEVs($\bar z_i$) alone is by no means guaranteed. In global
susy the $O(N_g)$ D terms   have the form  \bea D^a_{O(N_g)}&=&
Tr( \hat\phi^\dagger[T^a,\hat\phi]+   \hat{S}^\dagger[T^a,
 \hat{S}] ) +{\ovl D^a_X}\nnu
{ {{\bar{D}}^a_X}}&=&\sum_i\bar z_i^\dagger {{\cal{T }}^a} \bar
z_i \eea  where  ${\ovl D^a_X}$ is the contribution of the visible
sector fields,  and $T^a,{\cal T}^a$ the generators of $O(N_g)$ in
  the fundamental and generic representations.

Since we wish to use a flat direction of the hidden sector
superpotential to cancel the contribution ${\ovl{D^a_X}}$ of the
visible sector we  consider  the situation where $\phi_s,S_s$
become the trace modes of $O(N_g)$ symmetric representations
$\phi_{ab},S_{ab}$ so that the BM superpotential becomes \bea
W_H&=& Tr S(\mu_B \phi+ \sqrt{N_g} \lambda_B \phi^2) \eea
Generically \bea S&=& {\widehat{S}} +{\frac{1}{\sqrt{N_g}}} S_s \,
{{\cal I}_{N_g}} \qquad ;\qquad Tr {\hat {S}}=0\eea   ${{\cal
I}_{N_g}}$ is the $N_g$ dimensional unit matrix.

Before performing the case-wise analysis  consider the general
features of the minimization of the Bajc-Melfo potential arising
from the hidden sector potential $W_H$ alone when $N_g>1$. The BM
local minimum we find is characterized by an indeterminate VEV of
all the fields $S_s,\hat S$  in the symmetric multiplet $S$ while
the VEV of $\phi$ is fixed. The arguments given above and the
evaluations in the next section show that the VEV of the family
singlet in S is determined by Supergravity effects while the
others are fixed by minimizing the family  D-terms(which are of
$O(M_{GUT}^4))$ together with the leading corrections to them
namely the Supergravity induced soft mass terms ($O(m_{3/2}^2
M_{GUT}^2 $): all of which starts from considering the BM local
supersymmetry breaking minimum and its flat direction for $N_g>1$
together with the globally supersymmetric vevs $\bar z_i$ as the
background in which the determination of the sliding multiplet
vevs $S_s,\hat S$ are determined. We built up the argument for
this fixation stepwise in Section 2 and the novelty is now only
the determination of the $\hat S$ vevs.

 The F-terms arising from $N_g$ are
 \bea F_\phi &=& \mu_B S + \lambda \sqrt{N_g} ( S.\phi +\phi.S)
 \nnu
 F_S &=& \mu_B \phi + \lambda \sqrt{N_g} \phi.\phi \eea
 Clearly the associated global superpotential
 \bea V(\phi,S) &=& Tr(F_S^\dagger F_S +F_\phi^\dagger . F_\phi )\eea
 is semipositive definite. We search for the generalization of the BM
 local minimum by the ansatz \bea <\phi> ={\frac{1}{\sqrt{N_g}}}
  \bar\phi_s
{{\cal I}_{N_g}}\label{phigen}\eea We emphasize that this implies
that the gauge non singlet vev $\hat\phi\equiv 0$  for all $N_g$.
In fact this feature implies that  after the introduction of
supergravity corrections ${\cal{F}}_{\hat \phi}= 0$ (unlike
${\cal{F}}_{\hat S}$ which develops a value $O(m_{3/2} M_{GUT})$
and thus makes a $M_{GUT}/M_{Planck}$ suppressed contribution  to
the goldstino). With the ansatz (\ref{phigen})
 Then \bea <F_\phi>&=&<S> (\mu_B +   2\lambda
\bar\phi_s)\label{FSgen}\nnu
 <F_S>&=&<\bar\phi_s> (\mu_B + \lambda
\bar\phi_s){{\cal I}_{N_g}}\eea so that $\phi_s=-\mu_B/(2\lambda)
$ still ensures that $<F_\phi >,<F_{\hat S}>$ vanish  provided
$<\phi>$ has the form in (\ref{phigen}). Moreover the fluctuation
$\tilde \phi$ around this VEV receives a positive definite mass
squared \bea V=
 4|\lambda <S_s>|^2\, Tr \,\tilde\phi^\dagger \tilde\phi +....\eea
 from the VEV of the singlet in $S_s$ which is undetermined by the BM potential but
 fixed by Supergravity corrections to be be of  order $M_p$  as shown
 above. These contributions dominate all others. On the other hand \bea <F_S>=-\frac{\mu_B^2}{4
 \lambda}{\cal{I}}_{N_g} \eea breaks supersymmetry and causes
 corrections of order $\pm \mu_B^2$ to the  scalar masses of the
 $\phi$ supermultiplet. We again emphasize that the $F$ term of
 $\hat S$ vanishes for the BM  minimum for all $N_g$. The mass spectrum is
  discussed in detail in Section
 4.

\subsection{Determination of $S_\pm $  when  $N_g=2$ }
 For $N_g=2$ it
is convenient to use the isomorphism $O(2)\simeq U(1)$ and define
\bea S &=& {\frac{1}{2}}
\begin{pmatrix} \sqrt{2} S_s + S_{+} + S_{-}  &    i(S_{-}- S_{+})\cr
 i(S_{-}- S_{+}) & \sqrt{2} S_s -(  S_{+}+ S_{-})\cr \end {pmatrix}  \eea
and similarly for $\phi=[\phi_{ab}]$ where $S_\pm,S_s$ are
properly normalized fields so that $Tr S^\dagger S = S_+^\dagger
S_+ + S_{-}^\dagger S_{-} + S_s^\dagger S_s$ and $ Tr S \phi= S_s
\phi_s + S_{(+}\phi_{-)} $. In this notation an $O(2)$ vector
$\psi_a$ has charges $\pm 1/2 : \psi_{\pm 1/2}=(\psi_1\pm i
\psi_2)/\sqrt{2}$. The superpotential becomes \bea W_H&=&
  \mu_B (S_s  \phi_s  + S_{(+}\phi_{-)})+ \lambda_B
  ( S_s \phi_s^2 + 2 S_s\phi_+\phi_{-} + 2 \phi_s S_{(+}\phi_{-)}) \label{WH2}  \eea
and it is easy to check that   $\bar{\phi}_s=-\mu_B/2\lambda_B,
\bar{\phi}_{\pm }=0$ solves the $F$-term conditions for $W_H$ (all
F -terms vanish except   that ${\overline{\partial_{S_s}
W}}=\theta$ as before). This verifies that the general form of the
the vevs of $\phi,F_S$ is indeed that given in
(\ref{phigen},\ref{FSgen}).  The fields $S_{s,\pm}$ remain
undetermined at this local minimum and the superpotential reduces
to the one for $N_g=1$ when the fields that \emph{are} determined
are equated to their values at the local minimum. Notice that the
vev $\bar \phi$ makes the coefficient of $S_{(+}\phi_{-)}$
\emph{vanish} so that there is no Dirac mass mixing the chiral
fermions in $S_{\pm},\phi_{\mp}$.

The calculation for the singlet goes through unchanged fixing
$S_s\sim M_p$ as in eqns.(\ref{phigen},\ref{FSgen}).  With this
VEV the $O(2)$ charged fermionic components of $\hat\phi_\pm$ get
masses $\sim \lambda_B \bar S_s$. Along with the GUT scale
breaking of the family symmetry the  large mass of the modes in
$\phi_{ab}$ is responsible for quelling the percolation of the
supersymmetry breaking  coded in $F_{S_s}$  and ensuring the
hidden sector is actually  hidden.  Since $\hat\phi_\pm$ have zero
VEVs at the Susy breaking minimum they do not mix with the family
symmetry gaugino. Since the charged S fields can   have VEVs  of
at most the  order of  the GUT scale $M_{GUT}<< M_P$ the potential
for their determination to leading order in $m_{3/2}$ is simply
the common scalar mass term generated by the Supergravity
potential and the $D$ term of the family symmetry.
     In the O(2) case the leading
order potential for the flat family charged $S$ directions is
  \bea V(S_+,S_{-})&=& m_{3/2}^2(|S_+|^2 + |S_{-}|^2) +
{\frac{g_f^2}{2}}  (|S_{+}|^2 -|S_{-}|^2 +\bar{D}_X)^2 \eea here
$\bar{D}_X=\sum q_i |\bar{z}_i|^2 $ where $q_i,\bar{z}_i$ are the
family symmetry charges of the  global supersymmetric VEVs in the
visible sector. Clearly, if $x=Sign[D_X]$  the minimum will occur
when \bea
  S_{-x} &=& \sqrt{|\bar{D}_X| - x {\frac{m_{3/2}^2}{g^2}}} \qquad ;
 \qquad   \bar{S}_{x}  =0 \eea
Notice that the shift in the $S_{-x}$ VEV relative to
$\sqrt{\bar{D}_X}$ is tiny  but the gravity induced mass term does
enforce the choice of which component of $\hat S$ gets a VEV. The
vacuum energy after minimization is $\sim m_{3/2}^2 |D_X|$ and can
be tuned to zero by a shift $\delta {\hat W_0} \sim m_{3/2}|D_X| <
m_{3/2} M_p^2$. Note that since $\bar F_{ \hat S} =0$ from
(\ref{WH2})(and $\bar F_{z_i}=0$ by our scenario) the supergravity
corrections to the global D-terms shown in (\ref{Vsgrav}) are
zero.

\subsection{Fixation of   S VEVs for  $N_g=3$} For $N_g=3$ we
define the components of $S_{ab}$ in terms of $T_3$ eigenfields to
be  \bea\left (
  \begin {array} {ccc} \frac {1} {6} \left (\sqrt {6} {S_ 0} + 3 i{S_ {-2}} - 3 i {S_ {+2}} +
 2 \sqrt {3} {S_s} \right) & \frac {1} {2}  ({S_{+2}} + {S_ {-2}}) & \frac{1}{2} ({{S_ {+}} + {S_{-}}})   \\
\frac {1} {2}   ({S_{-2}} + {S_{+2}}) & \frac {1}{6} \left (\sqrt
{6} {S_ 0} - 3 i{S_ {-2}} + 3 i{S_{+2}} + 2 \sqrt {3} {S_s}
\right) & \frac {1} {2} i ({S_ {+}} - {S_{-}})\\ \frac{1}{2}( {{S_
{+}} + {S_{-}}})   & \frac {1} {2} i ({S_{+}} - {S_{-}}) &
\frac{1}{\sqrt {3}} ({{S_s} - \sqrt {2} {S_ 0}})
    \end {array}\
   \right)
   \nonumber\eea
and similarly for $\phi=[\phi_{ab}] $.  $S_{s,0,\pm,\pm 2} $ are
properly normalized fields so that \bea Tr S^\dagger S &=&S_{+2
}^\dagger S_{+2} + S_{-2}^\dagger S_{-2} + S_+^\dagger S_+ +
S_{-}^\dagger S_{-} + S_s^\dagger S_s\nnu  Tr S \phi &=& S_s
\phi_s + S_{(+}\phi_{-)}  + S_{(+2}\phi_{-2)}\eea

In this case the explicit form of the BM superpotential
is((square)round brackets denote(anti-)symmetrization)  \bea
W_{BM} &=& \mu_B (S_s \phi_s + S_0 \phi_0 + S_{(+}\phi_{-)} +
S_{(+2}\phi_{-2)} )\nnu && +\lambda_B\big[( S_s (\phi_s^2 +
\phi_0^2 + 2 \phi{_-}\phi{_+} +2 \phi_{-2}\phi_{+2}) \nnu && + S_0
(-\phi_0^2/\sqrt{2}+ 2 \phi_s \phi_0 -   {\sqrt{2}}
\phi{_-}\phi{_+} +\sqrt{2} \phi_{-2}\phi_{+2})\nnu && + 2
S_{(+}\phi_{-)}\phi_s -{\frac{1}{\sqrt{2}}}S_{(+}\phi_{-)}\phi_0
+i\sqrt{3}( S_+\phi_{-2}\phi_+ - S_-\phi_-\phi_{+2})\nnu &&
+\sqrt{2} S_{(+2}\phi_{-2)}\phi_0 + 2 S_{(+2}\phi_{-2)}\phi_s +
{\frac{i}{2}}S_{[+2}\phi_{-2]}^2\big]\eea

Using this one may again verify that the vevs are of the form
given in eqns.(\ref{phigen},\ref{FSgen}).

 When $N_g=3$ the D-terms form a vector of $O(3)$ and it is
 advantageous to change to a basis(denoted by primes) where
  ${\bar{D'}}_X^a=\delta^a_3 { {|\bar D_X|}} $.   This is easily achieved by
  a  rotation in the 12 plane to rotate the
$\bar{D}_X^a$-vector into the 23 plane and then a 23 plane
rotation to make it
 point in the 3 direction :
 \bea
O &=& R_{23}[\theta_X] . R_{12}[{\frac{\pi}{2}}-\varphi_X] \nnu
\theta_X&=&ArcTan[\sqrt{\frac{V_1^2 + V_2^2}{V_3^2}}]\quad ;\qquad
\varphi_X  = ArcTan[ \frac {V_2}{V_1}] \eea  where $V^a=
\bar{D}_X^a$. The potential for the flat directions from $\hat S'$
is now\bea V[\hat S'] &=& m_{3/2}^2(|S_0'|^2 +|S_{+}'|^2
+|S_{-1}'|^2 + |S_{+2}'|^2 +|S_{-2}'|^2)\nnu &+& {\frac{g_f^2
}{2}}\{(|S_{+}'|^2 +2|S_{+2}'|^2 -|S_{-}'|^2 -2|S'_{-2}|^2
+(\bar{D}_X^3)')^2  \nnu && +  2
 Tr({S'}^\dagger [T_+,S']) Tr({S'}^\dagger [T_{-},S'])  \}\eea This
has a solution along the same lines as for the $N_g=2$ case.
Dropping primes one gets : \bea
 |\bar{S}_{-2 }| &=& \sqrt{ {\frac{|{\vec D_X}|}{2}} - {\frac{m_{3/2}^2
}{4 g_f^2}}}  \nnu   \bar{S}_{-,+,+2}  &=& 0\eea
\section{Mass Spectrum}
The spectrum from  the visible sector gauge interactions and
superpotential is sensible by assumption and worked out for
realistic cases in \cite{yukawon}. Our interest here is in the BM
fields and-where relevant- their mixing with visible sector fields
through gauge or supergravity terms. We shall confine ourselves to
working out the leading contributions and ignore the effects of
small shifts ($\bar z_i\rightarrow \bar z_i +\delta_i$ , where
$\delta_i$ of order $m_{3/2}$) in visible sector vevs which  we
assume (via the usual shift  $\delta_i=-(W'')^{-1}_{ij} \bar z^j
\kappa^2 {\ovl W}$)     maintain ${\cal{F}}_{z_i}=0$.  We first
consider the mass terms that arise in the globally supersymmetric
Lagrangian when the fields $\phi_s,z_i,S_s,\hat S$ are replaced by
their vevs. The supergravity corrections are necessarily
suppressed by the Planck Length and are therefore important only
if a field does not obtain a mass at one of the large scales
present $\{M_{P},M_{GUT},\sqrt{M_{P} M_{3/2}} \} >> m_{3/2}$. At
the global level  vevs of F terms do not couple to fermion fields
at all so that any Susy breaking contributions will be visible
only in the scalar sector. Since only $F_{S_s}\neq 0$ it follows
that it makes a  susy breaking mass contribution  of order $\bar
F_{S_s}\sim m_{3/2}M_P$ to the scalars that it couples to in the
cubic terms of the superpotential.

In the family gauge terms (we ignore the visible gauge sector
under which the BM fields are singlets) the (global Susy
preserving) $\hat S$ vevs mix the chiral fermion in the $\hat S$
multiplet with the visible sector chiral fermion combination that
couples to the family gauginos to form a Dirac multiplet. Since
$\hat S$ terms otherwise lack mass terms this leads to a residual
massless fermion from the orthogonal mode. This is explained more
explicitly below.

 In the supergravity Lagrangian only fermion
bilinears coupling   with nonvanishing ${\cal{F}}$ terms can can
possibly obtain mass terms sensitive to supersymmetry breaking.
Only ${\cal{F}}_{S_s}\sim m_{3/2} M_{Planck}$ can be non zero at
leading order although the ${\cal{F}}_{z_i,\hat S}$ will get
smaller contributions of order $m_{3/2} M_{GUT}$
\subsection{Mass spectrum for $N_g=1$}

In this case only gravity couples the visible and hidden sectors.
The corrections to the visible sector are the familiar soft
supersymmetry breaking supergravity corrections which are
insensitive to the detailed form of the hidden sector
superpotential or whether its Susy breaking  minimum  is local or
global. The VEV $ \bar{\phi}_s =-\mu_B/2
 \lambda_B$ implies that the field $S_s$ (which has no mass term $W=m_S
 S_s^2 +..$ )  does not get a mass by mixing with $\phi_s$ and  provides
  the Goldstino as expected, even though the
 Supersymmetry is broken by a local rather than global minimum.
The fermion masses  consist of the $\phi_s$ mode with mass $ 2
\lambda_B  \bar{S}_s $  and the gravitino with longitudinal mode
from the goldstino  ${\tilde S_s}$.

In the BM sector the \emph{only} fermion mass terms at the global
level are \bea-{\cal{L}} &=& \lambda_B{\bar S_s}
\tilde{\phi}_s{\tilde\phi}_s
={\frac{m_{\tilde\phi_s}}{2}}\tilde{\phi}_s\tilde{\phi}_s \eea
Mass terms due to the vev $\bar \phi_s$ cancel.
 The scalar mass spectrum consists of two real scalar modes
from $\phi_s$ with masses squared $|2 \lambda_B \bar{S}|^2 \pm
|\mu_B|^2/2 $ (notice the supersymmetry breaking splitting of the
scalars  from the chiral fermion's Majorana  mass $2 \lambda_B
\bar S_s$) and two real scalars from $S_s$ with masses squared
$m_{3/2}^2 ((7+4 \sqrt{3})\pm(5+3 \sqrt{3}))$. This completes the
discussion of the $N_g=1$ case apart from explicitly working out
the effective theory which will be of the usual supergravity type
(Global Susy GUT derived MSSM  plus soft Susy breaking terms).

\subsection{Mass spectrum for $N_g=2$}

  Consider the masses of the superfields $\hat\phi_{ab}$. The fermions   do not
couple directly to the Susy violating F-term.  Inserting the vevs
in the fermion bilinear terms obtained from the superpotential in
eqn.(\ref{WH2}) we get the following  fermion mass terms
\emph{only}( tildes here denote fermionic components) \bea
-{\cal{L}} &=&
 \lambda_B {\bar S }_s ({\tilde\phi}_s\phi_s + 2 {\tilde\phi}_+{\tilde\phi}_-) + 2 \lambda_B {\tilde\phi}_s
 {\tilde\phi}_{(-}{\bar S}_{+)} \eea

 Since ${\bar S }_s>> \bar S_\pm $ it is clear that the ${\tilde\phi}_s
 {\tilde\phi}_\pm$ mixing is far subdominant  to the diagonal Majorana and Dirac
  mass terms ${\tilde\phi}_s^2,{\tilde\phi}_+{\tilde\phi}_-$.  Since there is no
residual gauge group the chiral mass matrix is just
${\ovl{\partial_i\partial_j W}}$.  We have a symmetric $3 \times
3$ chiral mass matrix with rows and columns labelled by
 $\{\phi_s,\phi_+,\phi_{-}\}$ :
 \bea m_{\phi\phi} &=& \lambda_B {\ovl S_s} \begin{pmatrix} 1 & \epsilon \,\, & 0\cr
 \epsilon \,\, & 0 \,\, & 1\cr 0 \,\, & 1 \,\, & 0\cr\end{pmatrix}\eea
 where $\epsilon = \ovl S_{-x}/\ovl S_s \sim
 M_X/M_p$.  This can easily be diagonalized and gives large
 masses in the $GUT$ scale range for reasonable values $\lambda_B \sim 10^{-3}$
 to $10^{-1}$. In the scalar sector the $\phi$
 fields also receive mass-squared  contributions from $\lambda_B F_{S_s}\sim
 \mu_B^2$ besides the contributions shown for the fermions. However
 since $|\mu_B|<<M_X <<M_p$ these will lead to small splitting of
 the dominant contributions shared with the fermions. All this is
 a straightforward extension of the behavior of the BM $\phi$ for $N_g=1$.

With the  VEVs  given in Section 3 for $N_g=2$ and due to the lack
of any other mass term for $S_\pm$  it is easy to check that the
fermionic component of $S_{-x}$ mixes  with the visible sector
Higgsino  field combination that combines to form a Dirac
multiplet with the single family gaugino of the $O(2)$ gauge
symmetry because the visible sector Higgs fields carry symmetric
reppresentation of the family group. Thus the Dirac partner
$\Lambda $ for the $U(1)_f$ gaugino $\lambda_f $  with Dirac mass
\bea m_{\lambda_B} &=& \sqrt{2} g_f \sqrt{|\bar{S}_{-s}|^2 +\sum_i
q_i^2 |\bar z_i|^2} \eea is defined by  \bea  {\cal L} &=&
-m_{\lambda} \Lambda \lambda_g +H.c. \nnu \Lambda &=&
(\cos[\theta_S] \tilde{Z} - \sin[\theta_S]
\tilde{S}_{-x}){\frac{\sqrt{2} g_f}{m_{\lambda_B}}}\nnu
\tilde{Z}&=& \frac{\sum_i q_i\bar{z}_i^* \tilde{z}_i}{\sqrt{\sum_i
q_i^2 |\bar z_i|^2}} \nnu \tan\theta_S&=&
\frac{x|\bar{S}_{-x}|}{\sqrt{\sum_i q_i^2 |\bar z_i|^2}} \eea
\emph{However due to the absence of other mixing terms  for
$S_{\pm s}$  the orthogonal combination $\tilde  \chi$ and $S_x$
remain massless at leading order}. Here \bea \tilde{\chi} &=&
\cos[\theta_S] {\tilde{S}_{-x}} +\sin[\theta_S] \tilde{Z} \eea
  does not mix with the remaining massive modes since it is made of (SM neutral) zero
modes $\tilde S_{-x},\Lambda$  of the chiral mass matrices in the
hidden and visible sectors respectively. Shifts due to
supergravity are too small to shift the masses $
m_{S_+\phi_{-}}=m_{\phi_+S_{-}} =\mu_B + 2 \lam \bar{\phi}_s =0$
appreciably. So the leading contributions come from loop
effects(see Fig.1) (analogous to the one loop corrections to
neutrino masses induced by  soft susy
breaking\cite{susynumasscalc}) and have value \bea
m_{\hat{S}}^{1-loop} \simeq \frac{|\lambda_B|^2}{16\pi^2}
\frac{\bar F_S}{\bar S_s} \simeq \frac{|\lambda_B|^2}{16\pi^2}
\frac{m_{3/2}}{\sqrt{3}-1}\eea Thus these masses may be as small
as a few GeV even for large gravitino mass. Due to their weak
interactions with light modes they may be suitable candidates for
light Cold Dark matter.
\begin{figure}\includegraphics[scale=.65]{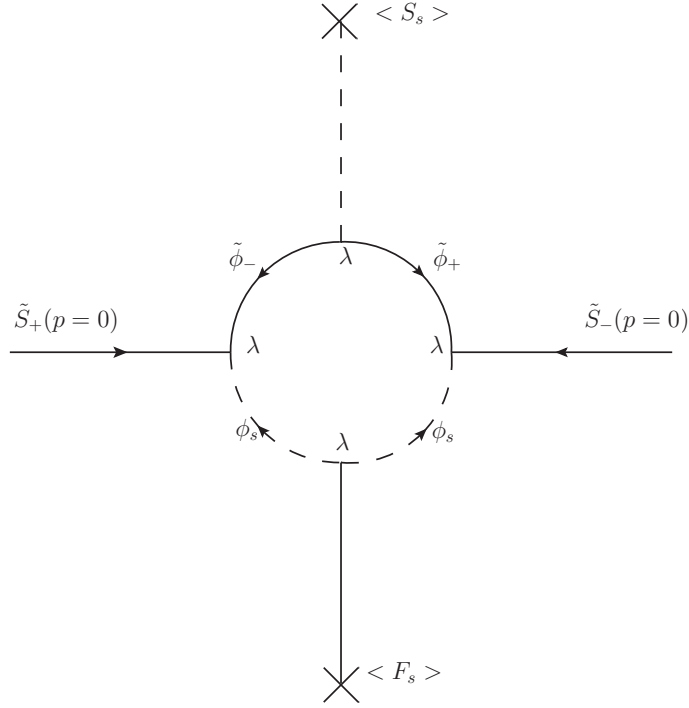}{\caption{Diagram for
fermionic  mass $m_{\tilde{S}}$}}\end{figure}
   The scalar partners  have masses \bea (m^2)_{S_x} &=& 2 m_{3/2}^2
\quad ;\qquad m^2_{Re(S_{-x})} = 2 g_f^2
 |D_X| -2 m_{3/2}^2 \eea Like $\tilde S_{-x}$  the hidden sector Goldstone scalar  component
$Im[S_{-x}]$ will again lead to a nearly massless mode since it
too will mix with the $O(2)$ Goldstone contributions from the
visible sector and hence the orthogonal mode (partner of the
pseudo-Goldstino $\tilde{\chi}$) will be massless before loop
effects.  The scalar pseudo-Goldstone mode $\chi$ which remains
massless even  after symmetry breaking due to pseudo-doubling of
goldstone modes(i.e independent breaking of the family symmetry in
the potentials from the unlinked Hidden and visible sector
superpotentials)  can also obtain loop induced mass from the
partner of the graph shown for the fermions, but it will remain
light. This is because the supersymmetry breaking F term couples
to other particles through $\phi_{ab}$ propagators
 whose large masses $\lambda_B <S_s>\sim M_X$ ensure that the effect of  one loop corrections
 is  a power  of   $F/M_X \sim m_{3/2}M_p/M_X$.
It is clear that since all the other modes have large masses much
greater than the gravitino mass and we have already taken into
account the dominant contributions to the fields that participate
in the Susy breaking, the supergravity corrections to their masses
are sub-dominant and of no immediate significance to warrant their
presentation  here.

\subsection{Mass spectrum for $N_g=3$}

The analysis for this case proceeds in a manner closely parallel
to the case of two generations. Once again all contributions from
the vev of $\phi_s$ cancel leaving for the BM fermion mass terms
\emph{only}  \bea -{\cal{L}} &=&
 \lambda_B {\bar S }_s ({\tilde\phi}_s{\tilde\phi}_s
+{\tilde\phi}_0{\tilde\phi}_0 + 2 {\tilde\phi}_+{\tilde\phi}_- + 2
{\tilde\phi}_{+2}{\tilde\phi}_{-2} )
  + 2 \lambda_B {\tilde\phi}_s {\tilde\phi}_{+2}{\bar S}_{-2} \eea

Once again we have diagonal Majorana/Dirac masses $2 \lambda_B
\bar S_s$ for all the components of $\phi$ and much smaller off
diagonal contribution from the vev of the non-singlet $\hat S$.
These will be accompanied by scalar masses of the same magnitude
up to small splitting $O(m_{3/2})$. Similarly the  scalars from
the $S_s$ multiplet will obtain  masses of order the gravitino
mass.

As for  the $N_g=2$ case fermionic partners of the $\hat S$
particles (except $S_{-2}$ but even so the analog of
$\tilde{\chi}$ will remain light via  the mechanism explained for
the $N_g=2$ case)   do not get mass to leading order. The Dirac
partners of the family symmetry gauginos in the primed basis(where
$\bar D^a\sim \delta^a_3$  are :\bea \Lambda^3 &=&
 \frac{-2{\bar{S'}}_{-2}^\dagger   \tilde{S'}_{-2} +
\sum_i q_i\bar z_i'^\dagger \tilde{z'}_i}{(4\ovl{S'}_{-2}^\dagger
 \ovl{S'}_{-2}+\sum_i
 \bar q_i^2 z_i'^*   \bar{z'}_i)^{1/2}} \nnu
  \Lambda^{\pm} &=&
 \frac{\sum_i \bar z_i'^\dagger {\cal{T}}^\pm \tilde{z'}_i}{( \sum_i
 \bar z_i'^\dagger {\cal{T}}^+  {\cal{T}}^-\bar{z'}_i)^{1/2}}  \eea
As before the fermion mode orthogonal to $\Lambda^3$ will remain
massless along with $\widetilde S_{0,\pm  ,  2}$.

The scalar masses are positive( apart from the Goldstone boson)
\bea M^2_{S_{0}}&=& m_{3/2}^2 \quad ;\qquad
M^2_{S_{-}}={\frac{1}{2}} m_{3/2}^2 \nnu
M^2_{S_{+}}&=&{\frac{3}{2}}  m_{3/2}^2 \quad ;\qquad
M^2_{S_{+2}}=2 m_{3/2}^2 \nnu M^2_{R_{-2}}&=&4 g_f^2|{\frac{{\vec
|D_X}|}{2}} -
 2  m_{3/2}^2   \quad ;\qquad
M^2_{I_{-2}}=0 \qquad\eea   In the above-since we are working in
the primed basis- we have exhibited only the  Goldstone
longitudinal component of the $O(3)$ gauge boson in the $3$
direction which comes from the field $S_{-2}$ : of course all
three $O(N_g)$ massive gauge bosons will get longitudinal mode
contributions from the visible sector (which we assume breaks the
family symmetry completely) which we have not discussed explicitly
here (see however \cite{yukawon} for more details in a concrete
model based on a realistic SO(10) GUT). Since only a linear
combination of $I_{-2}$ and a visible sector massless mode will be
eaten, we would again expect the orthogonal combination to remain
massless.

 Now  we will have five  light fermionic and one
 scalar degree  of freedom lighter   than the gravitino mass
 and with very weak couplings to ordinary matter.
  Again they will  pick up mass at one
loop due to supersymmetry breaking and  will remain as the fossils
of flavour and super symmetry breaking. Thus they are either
candidates for multicomponent relic Dark matter or must be removed
somehow.

From the above discussion, it is clear that the BM Susy breaking
Minkowski vacuum is  characterized by the presence of several
light fields coming from the supersymmetry breaking ($S_s$) and
from the $O(N_g)$ variant $\hat{S}$ fields over and above the
usual MSSM light fields in the visible sector. The  superfields
($\phi_{ab}$) are  superheavy  with masses of order  $\lambda_B
\bar{S}_s \sim M_X$. The effective low energy theory must be
written in terms of both sets of light fields by equating the
heavy fields to their VEVs and separating the light fields out via
an expansion in $m_{3/2}/M_p$. This calculation proceeds largely
like that in \cite{halllykkwein} but care must be taken due to the
additional gauge invariance active in both hidden and gauge
sectors : the main effects of  which  the present analysis sought
to clarify.

It bears mention that radiative effects in such systems may be
quite significant and have an important bearing on cosmological
behaviour\cite{dinemeshanfisch}. The loop corrected Kahler
potential is available up to two loops\cite{nibbetino}. We will
return to the study of the vacuum and cosmological predictions in
the theory with both gravity and radiative effects in a sequel.

 \section{Discussion }

In this paper we have shown  how, gravity mediation of Bajc-Melfo
type calculable   Supersymmetry breaking at metastable vacua in a
hidden sector containing fields variant under  a family gauge
symmetry, ensures the viability of  GUT and family symmetry
 breaking in the  visible sector of a ``Grand
Yukawonification ''\cite{yukawon} model. In such  models one aims
at generating the observed hierarchical fermion  flavour structure
from a gauged family symmetry  model with only generation blind
couplings. The special role of the BM supersymmetry breaking is
that it provides flat directions in both the   $O(N_g)$ singlet
and the gauge variant parts of a symmetric chiral supermultiplet
$S_{ab}=S_{ba}$.   Since it is very difficult to make the
contribution of the visible sector GUT fields to the $O(N_g)$
D-terms vanish,   the $\hat S$ flat direction performs the
invaluable function of cancelling this contribution \emph{without
disturbing the symmetry breaking in the visible sector}. The
determination of the viability of the ``Grand Yukawonification''
model then becomes a matter of searching the relatively  small
remaining parameter space for viable parameter sets that fit the
fermion data at $M_X$ while taking account of threshold
corrections at low and high scales and while respecting
constraints on crucial quantities like the proton
lifetime\cite{ag2,nmsgut,bstabhedge}. Note that in this approach
not only are the hard parameters of the visible sector
superpotential reduced by replacement of the flavoured parameters
by bland family symmetric ones  but also the soft supersymmetry
breaking parameters are determined by the two parameters of the
hidden sector superpotential and the Planck scale.

The structure used entails yet further stringent constraints since
the masslessness of the   moduli multiplets  $\hat{S}_{ab}$ before
supersymmetry breaking  implies  the existence of $N_g(N_g+1)/2
-1$ SM singlet fermions generically lighter than the gravitino
mass scale and possibly as light as a few GeV. In addition the
Polonyi mode $S_s$ may  also lead  to difficulties in the
cosmological scenario. Thus such modes can be both a boon and a
curse for familion GUT models. A boon because generic Susy GUT
models are hard put if asked to provide Susy WIMPs of mass   below
100 GeV as CDM candidates as suggested by the DAMA/LIBRA
experiment\cite{damalibra}. A curse because there are strong
constraints on the existence of such light moduli which normally
demand that their mass be rather large ($> 10$ TeV)
 due to the robust   cosmological(`Polonyi') problems
 arising from   decoupled modes with Planck scale VEVs\cite{modulicon}.
  In contrast to the simple Polonyi model and
  String moduli, the BM moduli  have explicit   couplings  to light fields through
family D-term mixing and loops. Moreover the MSGUT scenario
favours\cite{nmsgut,gutupend,bstabhedge}  large gravitino masses
$> 5-50 $ TeV. Thus  the Polonyi  and moduli problems may be
evaded. In any case the cosmology need be considered seriously
only after we have shown that the MSSM fermion spectrum is indeed
generated by the "Yukawonified" NMSGUT\cite{yukawon}.

 \section*{Acknowledgements} I am very grateful to Borut Bajc for
 numerous valuable discussions. I thank Charanjit K. Khosa for
 collaboration and Ila Garg for discussions and  help with preparing the
 manuscript.

\end{document}